\documentclass[letter]{jpsj2} 
\newcommand{\bS}{\mbox{\boldmath{$S$}}}

\title{Replicated Bethe Free Energy: 
A Variational Principle behind Survey Propagation}
\author{Yoshiyuki Kabashima}

\inst{Department of Computational Intelligence and Systems 
Science, \\
Tokyo Institute of Technology,\\
Yokohama 226-8502, Japan}

\abst{
 A scheme to provide various mean-field-type 
 approximation algorithms is presented by employing the 
 Bethe free energy formalism to a family of replicated systems
 in conjunction with analytical continuation with 
 respect to the number of replicas. 
 In the scheme, survey propagation (SP), which is an efficient 
 algorithm developed recently for analyzing 
 the microscopic properties of glassy states for 
 a fixed sample of disordered systems,  
 can be reproduced by assuming the simplest
 replica symmetry on stationary points of the 
 replicated Bethe free energy. 
 Belief propagation and generalized SP can 
 also be offered in the identical framework
 under assumptions of the highest 
 and broken replica symmetries, respectively. 
}
\kword{survey propagation, belief propagation, 
Bethe approximation, replica method, spin glasses}

\begin{document}
\maketitle
Recent research in cross-disciplinary fields involving 
statistical mechanics and information sciences has shown 
that methods and concepts from physics can be useful in 
the development and analysis of efficient algorithms for 
computing probabilities or solving combinational 
problems~\cite{Nishimori}. One of the most prominent 
examples in such research is the development of survey 
propagation (SP)~\cite{K-SAT}. 
This algorithm approximately evaluates the 
microscopic averages of dynamical variables 
in a feasible amount of time for a fixed sample of disordered 
systems utilizing the concept of replica symmetry breaking (RSB), 
which was discovered in the study of spin glasses~\cite{Beyond}. 
SP has been reported to give excellent results 
in studies on spin glass models~\cite{Montanari_Ricci} 
and in various combinatorial problems~\cite{K-SAT,coloring}, 
promoting further extension of the applicable 
range~\cite{zecchina}. It has also been shown that 
SP reproduces the one-step RSB (1RSB) solution of 
replica theory~\cite{Beyond} when applied to 
the mean-field-type spin glass models~\cite{MP}. 
This clarity of behavior is an important 
feature of the SP algorithm. However, 
the nature of solutions found by SP for general 
systems remains poorly understood, which may 
restrict the range of possible applications. 

The aim of the present Letter is to partially 
resolve this problem. More precisely, 
a family of approximation algorithms employing 
the Bethe free energy formalism~\cite{Kikuchi,CVM} 
is proposed for replicated systems. 
Analytically extending the proposed algorithms 
with respect to the number 
of replicas $x = 1,2, \ldots$ to $x \in {\bf R}$ 
under the simplest replica symmetric ansatz 
turns out to provide a general expression 
of SP, in which $x$ plays the role of a parameter specifying 
the size of a subgroup of replicas in 
the conventional 1RSB scheme~\cite{Beyond}. 
In this way, SP can only converge to a point 
offered by the analytical continuity of stationary 
points of the $x$-replicated Bethe free energy. 

As a basis for proposed algorithms, consider a joint distribution of $N$ dimensional state variables $\bS=(S_1,S_2,\ldots,S_N)$, as given by
\begin{eqnarray}
P(\bS)=Z^{-1}
\prod_{\mu=1}^M {\psi_\mu(\bS_\mu)}
\prod_{l=1}^N {\psi_l(S_l)}, 
\label{joint}
\end{eqnarray}
where $\psi_\mu(\bS_\mu)$ and $\psi_l(S_l)$ are termed 
the clique and local evidences, which are dependent on a certain subset of multiple components (clique) $\bS_\mu$ and a single component $S_l$, respectively, and $Z=\sum_{\bS}\prod_{\mu=1}^M {\psi_\mu(\bS_\mu)} \prod_{l=1}^N {\psi_l(S_l)}$ is the partition function. For simplicity, only the Ising spin systems $\bS=\{+1,-1\}^N$ are considered here, but extension to other cases is straightforward. In such systems, evaluation of the marginal probabilities, 
\begin{eqnarray}
P(S_l)=\sum_{\bS \backslash S_l} P(\bS), 
\label{marginals}
\end{eqnarray}
is in general computationally difficult, where $X \backslash Y$ denotes a subset of $X$ from which $Y$ is excluded. The development of computationally tractable algorithms achieving an accurate approximation of eq.~(\ref{marginals}) is therefore of great importance.

The Kullback-Leibler divergence (KLD), as given by
\begin{eqnarray}
{\rm KL}(Q|P)=\sum_{\bS}Q(\bS)\ln \frac{Q(\bS)}{P(\bS)}, 
\label{KL}
\end{eqnarray}
where $Q(\bS)$ is an arbitrary test distribution of $\bS$, offers a useful guideline for developing such an algorithm. As ${\rm KL}(Q|P)$ is always nonnegative and minimized to zero if and only if $Q(\bS)=P(\bS)$, the minimization of ${\rm KL}(Q|P)$ with respect to $Q(\bS)$ generally yields the correct distribution $Q(\bS)=P(\bS)$. Direct application of this algorithm, however, is not particularly useful for computing eq.~(\ref{marginals}) because $P(\bS)$ itself is in general not computationally tractable. Instead, applying this variational (minimization) principle of KLD or its approximation 
to a family of tractable test distributions 
systematically leads to a number of potentially 
effective approximation algorithms. 

The central idea of the present treatment is the application 
of this principle not to the original system but to 
a family of replicated systems, as follows: 
\begin{eqnarray}
P_x(\{\bS^a\})
=
\prod_{a=1}^x P(\bS^a),
\label{replica_system}
\end{eqnarray}
where $\{\bS^a\}$ is an abbreviation of the set 
of replicated systems $\{\bS^1,\bS^2,\ldots,\bS^x\}$. 
The abbreviations $\{S_l^a\}$ and $\{\bS_\mu^a\}$ are 
also used to represent 
$\{S_l^1,S_l^2,\ldots,S_l^x\}$ and 
$\{\bS_\mu^1,\bS_\mu^2,\ldots,\bS_\mu^x\}$, 
respectively. Equation~(\ref{replica_system}) 
yields the following equality regarding free energy:
\begin{eqnarray}
-\ln Z=
{\cal F}_x(Q)
-\frac{1}{x}{\rm KL}(Q|P_x), 
\label{KLn}
\end{eqnarray}
which holds for arbitrary $x=1,2,\ldots$ and 
the test distribution $Q(\{\bS^a\})$, 
where 
\begin{eqnarray}
{\cal F}_x(Q)=\frac{1}{x}
\sum_{\{\bS^a\}}
Q(\{\bS^a\})
\ln 
\frac
{Q(\{\bS^a\})}
{\prod_{a=1}^x\left (\prod_{\mu=1}^M \psi_\mu (\bS^a_\mu) 
\prod_{l=1}^N \psi_l (S_l^a)
\right )}. 
\label{variational_f_enery}
\end{eqnarray}
As the true free energy $-\ln Z$ is constant, 
this implies that the minimization of ${\rm KL}(Q|P_x)$ 
is equivalent to that of ${\cal F}_x(Q)$. 
This yields the correct distribution 
$Q(\{\bS^a\})=
P_x(\{\bS^a\})=\prod_{a=1}^x P(\bS^a)$. 
Since $\prod_{a=1}^x P(\bS^a)$ is simply 
a product of duplications of eq.~(\ref{joint}), 
the minimization of eq.~(\ref{KL}) 
for the original system and 
${\cal F}_x(Q)$ 
for the replicated system are equivalent. 
The introduction of the family of replicas 
therefore provides no benefits as long as 
the search for the optimal test distribution 
covers all feasible functional spaces. 
However, the replicated formalism can potentially provide 
better approximation accuracy than 
the original algorithm when the test distributions 
are limited to a tractable family or when 
the cost functions are somewhat approximated, 
as the tractable family or characteristics 
of the approximated function depend on the number of replicas $x$. 

To show this, we approximate 
${\cal F}_x(Q)$ 
by the Bethe free energy~\cite{Kikuchi,CVM} of the $x$-replicated system, as follows:
\begin{eqnarray}
{\cal F}_x (Q) &\simeq& {\cal F}_x^{\rm Bethe}
(\{b_{\mu}\},\{b_{l}\}) \cr
&=& 
\frac{1}{x}\sum_{\mu=1}^M
\sum_{\{\bS^a_\mu\}}
b_{\mu}(\{\bS_\mu^a\})\ln 
\frac{b_{\mu}(\{\bS_\mu^a\})}{
\prod_{a=1}^x 
(\psi_\mu(\bS_\mu^a)
\prod_{l\in{\cal L}(\mu)}\psi_l(S_l^a))} \cr
&+&\frac{1}{x}\sum_{l=1}^N (1-C_l)
\sum_{\{S_l^a\}}
b_l(\{S_l^a\})\ln 
\frac{b_l(\{S_l^a\})}{
\prod_{a=1}^x \psi_l(S_l^a)}, 
\label{replica_bethe}
\end{eqnarray}
where 
${\cal L}(\mu)$ is the set of elements that directly relate
to clique $\mu$ and 
$C_l$ is the number of cliques to which element $S_l$ is directly related. The test distributions $b_\mu(\{\bS_\mu^a\})$ and $b_l(\{S_l^a\})$ are termed beliefs 
which approximate the marginals $\sum_{\{\bS^a\} \backslash \{\bS_\mu^a\}} P_x(\{\bS^a\})$ and $\sum_{\{\bS^a\} \backslash \{S_l^a\}} P_x(\{\bS^a\})$, respectively, when ${\cal F}_x^{\rm Bethe}$ is extremized. Note that since both of these marginals are reduced from the identical distribution $P_x(\{\bS^a\})$, the reducibility condition
\begin{eqnarray}
\sum_{\{\bS_\mu^a\} \backslash \{S_l^a \}}
b_\mu(\{\bS_\mu^a\})
=b_l(\{S_l^a\}), 
\label{reducibility}
\end{eqnarray}
must hold when $S_l$ is an element of $\bS_\mu$. 

If the variable dependence in eq.~(\ref{joint}) is represented by a cycle-free graph, eq.~(\ref{replica_bethe}) under this constraint agrees exactly with 
${\cal F}_x(Q)$ 
for the test distribution $Q(\{\bS^a\})=
\frac{\prod_{\mu=1}^M b_\mu(\{\bS_\mu^a\})}
{\prod_{l=1}^N b_l^{C_l-1}(\{S_l^a\})}$, in which case extremizing eq.~(\ref{replica_bethe}) leads to the exact assessment of eq.~(\ref{marginals})~\cite{CVM,Lauritzen}. Unfortunately, 
${\cal F}_x(Q)$ and ${\cal F}_x^{\rm Bethe}(\{b_\mu\},\{b_l\})$ 
do not accord in general. However, this indicates that the stationary point of eq.~(\ref{replica_bethe}) may provide a good approximation when the influence of the cycles can be regarded as weak. 


Extremizing eq.~(\ref{replica_bethe}) with respect to the beliefs, adding the terms $\sum_{\{S_l^a\}}\lambda_{\mu l}(\{S_l^a\}) 
\left (
\sum_{\{\bS_\mu^a\} \backslash \{S_l^a \}}
b_\mu(\{\bS_\mu^a\})
-b_l(\{S_l^a\})
\right )$, 
where $\lambda_{\mu l}(\{S_l^a\})$ are the Lagrange multipliers imposing constraint (\ref{reducibility}), yields
\begin{eqnarray}
b_\mu(\{\bS_\mu^a\})
&=&\frac{
\prod_{a=1}^x \psi_\mu(\bS_\mu^a) 
\prod_{l \in {\cal L}(\mu) }
(e^{-\lambda_{\mu l}(\{S_l^a\})}
\prod_{a=1}^x \psi_l(S_l^a))}
{\sum_{\{\bS_\mu^a\}}
\prod_{a=1}^x \psi_\mu(\bS_\mu^a) 
\prod_{l \in {\cal L}(\mu)}
( e^{-\lambda_{\mu l}(\{S_l^a\})}
\prod_{a=1}^x \psi_l(S_l^a))}, 
\label{lagrange_belief1} \\
b_l(\{S_l^a\})
&=&\frac{
\prod_{a=1}^x \psi_l(S_l^a)
\prod_{\mu \in {\cal M}(l)}
e^{-\frac{\lambda_{\mu l}(\{S_l^a\})}{C_l-1}}}
{\sum_{\{S_l^a\}}
\prod_{a=1}^x \psi_l(S_l^a)
\prod_{\mu \in {\cal M}(l)}
e^{-\frac{\lambda_{\mu l}(\{S_l^a\})}{C_l-1}}}. 
\label{lagrange_belief2} 
\end{eqnarray}
Then, inserting eqs.~(\ref{lagrange_belief1}) and (\ref{lagrange_belief2}) into eq.~(\ref{reducibility}) affords the stationary point condition of the Lagrange multipliers, which can be read as
\begin{eqnarray}
e^{-\hat{\lambda}_{\mu l}(\{S_l^a\})}
&\propto&
\sum_{\{\bS_\mu^a\}  \backslash \{S_l^a\}}
\prod_{a=1}^x \psi_\mu(\bS_\mu^a) 
\prod_{j  \in {\cal L} (\mu)\backslash  l}
(e^{-\lambda_{\mu j}(\{S_j^a\})}
\prod_{a=1}^x \psi_j(S_j^a)), 
\label{horizontal} \\
e^{-\lambda_{\mu l}(\{S_l^a\})}
& \propto  &
\prod_{\nu \in {\cal M}(l) \backslash \mu}
e^{-\hat{\lambda}_{\nu l}(\{S_l^a\})},
\label{vertical}
\end{eqnarray}
where 
${\cal M}(l)$ denotes 
the sets of cliques that directly 
relate to clique elements $l$
and $\hat{\lambda}_{\mu l}
(\{S_l^a\})=
\frac{1}{C_l-1}\sum_{\nu \in {\cal M}(l)}
\lambda_{\nu l}(\{S_l^a\})-\lambda_{\mu l}(\{S_l^a\})$.

A remarkable property of these replicated systems is the invariance of stationary conditions (\ref{horizontal}) and (\ref{vertical}) under the permutation of replica indices $a=1,2,\ldots,x$. This naturally leads to the assumption that the stationary point possesses the same symmetry. In the case of Ising spin systems $\bS =\{+1,-1\}^N$, this can be represented as the simplest replica symmetric (RS) ansatz on the Lagrange multipliers,
\begin{eqnarray}
&&
e^{-\lambda_{\mu l}
(\{S_l^a\}) }
\prod_{a=1}^x
\psi_l(S_l^a) \propto 
\int dh \pi_{l \to \mu}
(h) \frac{e^{h \sum_{a=1}^x S_l^a}}
{(2 \cosh (h))^x},   
\label{replace1}\\
&&e^{-\hat{\lambda}_{\mu l}
(\{S_l^a\}) }
\propto 
\int d\hat{h} \hat{\pi}_{\mu \to l}
(\hat{h}) \frac{e^{\hat{h}\sum_{a=1}^x S_l^a}}
{(2 \cosh (\hat{h}))^x  },
\label{replace2}
\end{eqnarray}
without a loss of generality~\cite{Monasson}, where $\pi_{l \to \mu}(h)$ and $\hat{\pi}_{\mu \to l}(\hat{h})$ are distributions that absorb the degree of freedom of the Lagrange multipliers. 
Inserting these multipliers into eqs.~(\ref{horizontal}) 
and (\ref{vertical})
provides a couple of saddle point equations
in which $\hat{\pi}_{\mu \to l}(\hat{h})$ 
and $\pi_{l \to \mu}(h)$ 
are solved explicitly for given 
$\pi_{l \to \mu}(h)$ and 
$\hat{\pi}_{\mu \to l}(\hat{h})$, respectively. 
The natural iteration of the resultant equations
yields a family of algorithms 
that are parameterized by the number 
of replicas $x=1,2,\ldots$, 
\begin{eqnarray}
&&\hat{\pi}^{t+1}_{\mu \to l}(\hat{h})
\!\propto  \!\!
\int \!\!\!\!\!\! \prod_{j  \in  {\cal L}(\mu)\backslash  l} 
\!\!\!\!\!\!
dh_j \pi^t_{j \to \mu}(h_j) 
\left (\!\!\sum_{\bS_\mu}\psi_\mu(\bS_\mu) \!\!\!\! 
\prod_{j \in {\cal L}(\mu ) \backslash l}\!\!\!\! 
\frac{e^{h_jS_j}}{2\cosh (h_j)} \!\! \right )^x
\!\!\! 
\delta\! 
\left (\hat{h}\!-\!\hat{h}(\{h_{j\in {\cal L}(\mu)\backslash l}
 \}) \right ), 
\label{SP1}\\
&&\pi^t_{l \to \mu}(h)
\! \propto\!
 (2 \cosh (h))^x 
\int \prod_{\nu \in {\cal M}(l) 
\backslash \mu}
\frac{d\hat{h}_\nu \hat{\pi}^t_{\nu \to l}(\hat{h}_\nu)}{
(2 \cosh (\hat{h_\nu}))^x }
\delta (h-h_l^0-\sum_{\nu \in {\cal M}(l) \backslash \mu}
\hat{h}_\nu ), 
\label{SP2}
\end{eqnarray}
where $t$ represents the number of iterations, 
$\hat{h}(\{h_{j\in {\cal L}(\mu)\backslash l}
 \})=\tanh^{-1}\left (
\frac{\sum_{\bS_\mu}S_l \psi_\mu(\bS_\mu) 
\prod_{j \in {\cal L}(\mu) \backslash l}
\left (\frac{e^{h_jS_j}}{2
\cosh (h_j)} \right )}
{\sum_{\bS_\mu} \psi_\mu(\bS_\mu) 
\prod_{j \in {\cal L}(\mu) \backslash l}
\left (\frac{e^{h_jS_j}}{2\cosh (h_j)} \right )} \right )
$
and $h_l^0=\tanh^{-1}
\left (\frac{\sum_{S_l}S_l \psi_l(S_l)}{\sum_{S_l}\psi_l(S_l)}
\right )$. 

There are two features to note here. First, the beliefs can be evaluated by inserting eq.~(\ref{replace1}) into eqs.~(\ref{lagrange_belief1}) and (\ref{lagrange_belief2}) using the stationary solution of eqs.~(\ref{SP1}) and (\ref{SP2}). Specifically, eq.~(\ref{lagrange_belief2}) can be assessed as 
$
b_l(\{S_l^a\}) =
\int dh \rho_l(h) \frac{e^{h\sum_{a=1}^x S_l^a}}{(2 \cosh (h))^x}, 
$
yielding the following approximation of eq.~(\ref{marginals}):
$P(S_l)\simeq 
\int dh \rho_l(h) \frac{e^{hS_l}}{2 \cosh (h_l)}, 
$
where 
\begin{eqnarray}
\rho_l(h)=\frac{(2 \cosh (h))^x \int 
\prod_{\mu \in {\cal M}(l)}
\frac{d\hat{h}_\mu \hat{\pi}_{\mu \to l}(\hat{h}_\mu)}{
(2\cosh (\hat{h}_\mu))^x} 
\delta(h-h_l^0-\sum_{\mu \in {\cal M}(l) }
\hat{h}_\mu) }
{ \int 
\prod_{\mu \in {\cal M}(l)}
\frac{d\hat{h}_\mu \hat{\pi}_{\mu \to l}(\hat{h}_\mu)}{
(2\cosh (\hat{h}_\mu))^x} 
(2 \cosh (h_l^0+ \sum_{\mu \in {\cal M}(l) }
\hat{h}_\mu))^x}. 
\label{rho}
\end{eqnarray}
As the functional updates of eqs.~(\ref{SP1}) and (\ref{SP2}) 
can be approximated by the Monte Carlo method on a practical 
time scale, these equations constitute a tractable 
algorithm for approximating the marginal 
probabilities (\ref{marginals}). 

Second, although $x=1,2,\ldots$ has been assumed, 
the expressions of eqs.~(\ref{SP1})-(\ref{rho}) 
can be analytically continued to $x \in {\bf R}$. 
In addition, the extremized values of the 
replicated Bethe free energy 
${\Phi}_x^{\rm Bethe}=\mathop{\rm Extr}_{
\{\{b_\mu\},\{b_l\}\} }
\left \{{\cal F}_x^{\rm Bethe} (\{b_\mu\},\{b_l\})
\right \}$
can also be readily extended to $x\in {\rm R}$ as follows:
\begin{eqnarray}
&&{ \Phi}_x^{\rm Bethe}=-\frac{1}{x}
\sum_{\mu=1}^M \ln 
\left [
\int \prod_{l \in {\cal L}(\mu)}
dh_l \pi_{l \to \mu}(h_l)
\left (\sum_{\bS_\mu}\psi_\mu(\bS_\mu)
\prod_{l \in {\cal L}(\mu)}
\left (\frac{e^{h_lS_l}}{2\cosh (h_l)}
\right ) \right )^x 
\right ] \cr 
&&-\frac{1}{x}
\sum_{l=1}^N (1-C_l) \ln 
\left [\int 
\prod_{\mu \in {\cal M}(l)}
d\hat{h}_\mu \hat{\pi}_{\mu \to l}(\hat{h}_\mu)
\left (
\sum_{S_l}
\psi_l(S_l)
\prod_{\mu \in {\cal M}(l)}
\left (\frac{e^{\hat{h}_\mu S_l}}{2\cosh (\hat{h}_\mu)}
\right )
\right ) ^x \right ], 
\label{free_extremum}
\end{eqnarray}
where $\pi_{l \to \mu}(h)$ and $\hat{\pi}_{\mu \to l}(\hat{h})$
are the stationary solutions of eqs. (\ref{SP1}) and (\ref{SP2}). 
Analytical extension of the variational functional 
(\ref{replica_bethe}), however,  is nontrivial. 
Nevertheless, the analytically continued expressions 
of eqs.~(\ref{SP1}) and (\ref{SP2}) eventually 
lead to a general expression of SP 
in which $\pi_{l \to \mu}^t(h)$ and 
$\hat{\pi}_{\mu \to l}^t(\hat{h})$ are termed surveys,  
which is the main result of the present treatment. 
In this relation, $x$ serves as the RSB parameter, 
which is introduced to represent the number of 
replicas in a subgroup in the conventional 1RSB formulation. 
For example, SP for the K-SAT problem of finite temperature, 
corresponding to eqs. (C1) and (C2) in 
ref.~\citen{Montanari_parisi_Ricci}, 
can be derived from eqs.~(\ref{SP2}) and (\ref{SP1}) by setting 
$x \to m$, $\psi_\mu(\bS_\mu)=e^{-2 \beta 
\prod_{l \in {\cal L}(\mu)} 
\frac{1-J_l^\mu S_l}{2}}$ and 
$\psi_l(S_l)=1$ in conjunction with rescaling fields as 
$\frac{h}{\beta} \to h$ and $\frac{\hat{h}}{\beta}
\to \hat{h}$, where $\beta$ is the inverse temperature and $J_l^\mu \in \{+1,-1\}$ are randomly predetermined constants. The derivation of SP for the famous Sherrington-Kirkpatrick (SK) model~\cite{SK} is shown briefly in the appendix. 

It has been reported that extremizing eq.~(\ref{free_extremum}) with respect to the RSB parameter $x$ 
provides a reasonable description of 
equilibrium states in the conventional replica theory~\cite{MP}. 
However, it may be difficult to deduce such 
a principle for the determination of $x$ using only 
the framework presented here. The derivation 
is complicated by the dependence of the approximation 
accuracy on $x$, which is generally related 
to the specific characteristics of the  
target system in a nontrivial manner, 
although certain extra constraints related 
to the concept of thermodynamic limits 
(e.g., homogeneity~\cite{Janis}) may 
be utilized for problems in the physics literature. 

The formalism presented here is related to other algorithms as follows. Equations~(\ref{SP1}) and (\ref{SP2}) always have special solutions of the forms $\pi^t_{l \to \mu}(h)=\delta(h-h_{l \to \mu}^t)$ and $\hat{\pi}_{\mu \to l}^t(\hat{h})=\delta(\hat{h}-\hat{h}_{\mu \to l}^t)$,
independently of $x$. 
The parameters $h_{l \to \mu}^t$ and 
$\hat{h}_{\mu \to l}^t$ are updated by
$
\hat{h}_{\mu \to l}^{t+1}=\hat{h}(\{
h^t_{j \in {\cal L}(\mu) \backslash l} \})$, 
$h_{l \to \mu}^t=h_l^0+\sum_{\nu \in {\cal M}(l)\backslash \mu}
\hat{h}_{\nu \to l}^t. $
Notice that these forms are an expression of the belief propagation (BP)~\cite{BP,BPSG} for eq.~(\ref{joint}), the fixed point of which is usually linked to the variational condition of the conventional Bethe free energy ${\cal F}^{\rm Bethe}={\cal F}_{x=1}^{\rm Bethe}$~\cite{Yedidia}. In the current formalism, on the other hand, BP can be characterized as the solution of the highest symmetry 
obtained assuming the Lagrange multipliers of 
the limited form 
$e^{-\lambda_{\mu l}(\{S_l^a\})} 
\prod_{a=1}^x \psi_l(S_l^a)
\propto \frac{e^{h_{l \to \mu}\sum_{a=1}^x S_l^a}}{
(2 \cosh (h_{l \to \mu}))^x}$, 
$e^{-\hat{\lambda}_{\mu l}(\{S_l^a\})} 
\propto \frac{e^{\hat{h}_{\mu \to l}\sum_{a=1}^x S_l^a}}{
(2 \cosh (\hat{h}_{\mu \to l}))^x}$, which exist for arbitrary RSB parameter $x \in {\bf R}$. This relationship between BP and the assumption of eqs.~(\ref{replace1}) and (\ref{replace2}) is analogous to that for the RS solution under the 1RSB ansatz in the conventional replica theory employed in spin glass research. 

The present algorithm can also be extended to the 
reduction of replica symmetry. For example, introducing the 1RSB 
ansatz~\cite{Monasson} on the Lagrange multipliers, we obtain 
\begin{eqnarray}
e^{-\lambda_{\mu l}(\{S_l^a\})}
\prod_{a=1}^x \psi_l(S_l^a)
&\propto& 
\int {\cal D}{\pi} { \Pi}_{l \to \mu}[\pi]
\prod_{\alpha=1}^{\frac{x}{m}} 
\left (\int dh^\alpha 
\pi(h^\alpha)
\frac{e^{h^\alpha 
\sum_{a \in {\cal I}(\alpha)} S_l^{ a}}}{
(2 \cosh (h^\alpha))^m} \right ), \\
e^{-\hat{\lambda}_{\mu l}(\{S_l^a\})} 
&\propto& 
\int {\cal D}{\hat{\pi}} {\hat{ \Pi}_{\mu \to l}}
[\hat{\pi}]
\prod_{\alpha=1}^{\frac{x}{m}} 
\left (\int d\hat{h}^\alpha 
\hat{\pi}(\hat{h}^\alpha)
\frac{e^{\hat{h}^\alpha 
\sum_{\alpha \in {\cal I}(\alpha)} S_l^{a}}}{
(2 \cosh (\hat{h}^\alpha))^m} \right ),
\end{eqnarray}
where ${\cal I}(\alpha)$ ($\alpha =1,2,\ldots,m)$ 
denotes subsets of replica indices of an equal size $m$
while ${\cal D}{\pi} { \Pi}_{l \to \mu}[\pi]$ and 
${\cal D}{\hat{\pi}} \hat{ \Pi}_{\mu \to l}[\hat{\pi}]$ 
represent the variational measures of 
distributions $\pi(h)$ and $\hat{\pi}(\hat{h})$, respectively. 
Analytically continuing this yields the conventional 
two-step RSB solution in the case of the SK model.
This indicates that generalizing this provides an algorithm 
corresponding to the $r+1$-step RSB solution 
of conventional replica analysis under 
the assumption that the Lagrange multipliers 
are of the form of the $r$-step RSB.  

In summary, a framework for the construction of 
a family of mean-field-type approximation algorithms 
was derived by introducing the Bethe free energy 
formalism for $x$-replicated systems. 
Analytically continuing the algorithm 
obtained for $x=1,2,\ldots$ to 
$x \in {\bf R}$ under the replica symmetric ansatz 
leads to a general expression of survey propagation 
for such systems (eq.~(\ref{joint})), 
in which $x$ plays the role of a replica 
symmetry breaking parameter in the 1RSB solution 
of conventional replica analysis. 
Belief propagation and generalized survey propagation 
can be reproduced from an identical variational 
functional (eq.~(\ref{replica_bethe})) corresponding 
to various levels of RSB ansatz on the replica symmetry 
of the Lagrange multipliers. 
This may be useful for clarifying the relationships 
between solutions obtained using these various algorithms. 

Although the focus here was on algorithms derived 
from natural iteration of eqs.~(\ref{horizontal})
and (\ref{vertical}), a range of schemes 
for the minimization of the Bethe free energy are 
available~\cite{CCCP,Komiya}. Furthermore, 
it is known that the Bethe free energy formalism 
itself can be generalized to a more advanced scheme 
as the cluster variation method~\cite{CVM}. 
Extending the present results to such schemes 
is a target of future work. 

\section*{Acknowledgment}
This work was supported in part by Grants-in-Aid
(Nos.~14084206 and~17340116) from MEXT/JSPS, Japan. 
Useful discussion with A. Hatabu is acknowledged. 
\appendix

\section{SP for the SK Model}
The Sherrington-Kirkpatrick (SK) model is an Ising spin system characterized by the Hamiltonian ${\cal H}(\bS)=
-\sum_{l>k} J_{lk}S_lS_k-\sum_{l=1}^N H_l S_l$, where 
$J_{lk}$ is sampled from an identical 
normal distribution ${\cal N}(J_0/N,J^2/N)$
independently of the unordered pair  
$\left \langle lk \right \rangle$,
and $H_l$ represents an external field. 
Identifying 
$\left \langle lk \right \rangle$ with $\mu$ 
leads to $\psi_\mu(\bS_\mu)=e^{\beta J_{lk}S_lS_k}$
and $\psi_l(S_l)=e^{\beta H_l S_l}$.

Two distinct properties of this system are the 
denseness of connectivity and the weakness of each coupling 
constant. Let $\hat{\pi}^t_{\mu \to l}
(\hat{h})$ be characterized by the two moments
$\int d \hat{h} \hat{\pi}^t_{\mu \to l}(\hat{h}) 
\hat{h}=a_{\mu \to l}^t$
and $\int d \hat{h} \hat{\pi}^t_{\mu \to l}(\hat{h}) 
(\hat{h}-a_{\mu \to l}^t)^2=v_{\mu \to l}^t$. 
This representation allows 
eqs.~(\ref{SP2}) and (\ref{rho}) 
to be expressed in terms of the $t$th update as
\begin{eqnarray}
\pi^t_{l \to \mu}(h) \simeq 
\frac{(2 \cosh (h))^x 
e^{-\frac{(h-\phi_{l \to \mu}^t)^2}
{2 \Delta_{l \to \mu}^t}}}{
\int 
d h (2 \cosh (h))^x 
e^{-\frac{(h-\phi_{l \to \mu}^t)^2}
{2 \Delta_{l \to \mu}^t}}}, 
\label{SKpi}
\end{eqnarray}
and 
\begin{eqnarray}
\rho_l^t(h)\simeq 
\frac{(2 \cosh (h))^x 
e^{
-\frac{(h-\phi_{l}^t)^2}
{2 \Delta_{l }^t} }}{
\int 
d h (2 \cosh (h))^x 
e^{-\frac{(h-\phi_{l }^t)^2}
{2 \Delta_{l }^t} }}, 
\label{SKrho}
\end{eqnarray}
where 
$
\phi_{l \to \mu}^t=\beta H_l+
\sum_{\nu \in {\cal M}(l) \backslash \mu}
a_{\nu \to l}^t
$,
$\Delta_{l \to \mu}^t=
\sum_{\nu \in {\cal M}(l) \backslash \mu}
v_{\nu \to l}^t$, 
$\phi_{l }^t=\beta H_l+
\sum_{\mu \in {\cal M}(l) }
a_{\mu \to l}^t$ and 
$\Delta_l^t=
\sum_{\mu\in {\cal M}(l)} v_{\mu \to l}^t$~\cite{Ketsuago}. 
In turn, let us introduce 
$\int dh 
\pi_{l \to \mu}^t (h) \tanh(h)
=m_{l \to \mu}^t$, 
$\int dh 
\pi_{l \to \mu}^t (h) \tanh^2(h)
=M_{l \to \mu}^t$, 
$\int dh 
\rho_{l }^t (h) \tanh(h)
=m_{l}^t$
and 
$\int dh 
\rho_{l }^t (h) \tanh^2(h)
=M_{l}^t$.
Here, $m_l^t$ is the estimated local magnetization at the $t$th update. The weakness of each coupling constant in conjunction with eq.~(\ref{SP1}) indicates that $a_{\mu \to l}^t$ and 
$v_{\mu \to l}^t$ are updated as
$a_{\mu \to l}^{t+1} \simeq \beta J_{\mu} m_{k \to \mu}^t$, 
$v_{\mu \to l}^{t+1} \simeq \beta^2 J_{\mu}^2
(M_{k \to \mu}^t-(m_{k \to \mu}^t)^2)$. 
Employing the law of large numbers, this 
implies that variances $\Delta_{l \to \mu}^{t}$ and 
$\Delta_{l }^{t}$ are updated independently 
of the pairs of indices $(l\mu)$ as 
$\Delta_{l \to \mu}^{t+1} 
\simeq N^{-1}\beta^2 J^2
\sum_{\nu \in {\cal M}(l) \backslash \mu}
(M_{l \to \nu}^{t}-(m_{l \to \nu}^{t})^2)
\simeq 
N^{-1}\beta^2 J^2
\sum_{\mu \in {\cal M}(l) }
(M_{l }^{t}-(m_{l }^{t})^2)
\simeq \Delta_l^{t+1}
\simeq \beta^2J^2(Q_1^t-Q_0^t)$, 
where $Q_1^t=N^{-1}\sum_{l=1}^N M_l^t$ and 
$Q_0^t=N^{-1}\sum_{l=1}^N (m_l^t)^2$. 
This also indicates that 
$m_{l \to \mu}^t$ and $m_l^t$ are 
related by
$
m_{l \to \mu}^t 
\simeq
m_l^t-\beta J_\mu (1-Q_1^t+x(Q_1^t-Q_0^t))m_k^{t-1}
$
via the Taylor expansion. 
Inserting these results into eqs.~(\ref{SKpi}) and (\ref{SKrho}) leads to the following expression of SP for the SK model: 
\begin{eqnarray}
\phi_l^{t+1}
\!\!\!& =& \!\!\!\beta H_l+\sum_{k \ne l}
\beta J_{lk} m_k^t
-
\Gamma^t m_l^{t-1}, 
\label{fastSP1} \\
m_l^t \!\!\!&=&\!\!\!\frac{\int dh 
(2\cosh (h))^x 
e^{-\frac{(h-\phi_l^t)^2}{2\beta^2J^2(Q_1^t-Q_0^t)} 
}\tanh(h)}
{\int dh 
(2\cosh (h))^x 
e^{
-\frac{(h-\phi_l^t)^2}{2\beta^2J^2(Q_1^t-Q_0^t)} 
}}, 
\label{fastSP2}
\end{eqnarray}
where $\Gamma^t=\beta^2 J^2(1-Q_1^t+x(Q_1^t-Q_0^t))$, and $\sum_{\mu \in {\cal M}(l)}J_\mu^2=
\sum_{k \ne l}J_{lk}^2$ is replaced with 
its expectation $(N-1)N^{-1}J^2\simeq J^2$ 
using the law of large numbers. 

Notice that the fixed point condition 
of eqs.~(\ref{fastSP1}) and (\ref{fastSP2}) 
is in agreement with the 1RSB analogue to 
the Thouless-Anderson-Palmer equation 
of the SK model~\cite{Beyond,TAP}. 
This demonstrates the consistency of SP with the 1RSB solution 
in conventional replica theory.

\end{document}